\documentclass{rmaa}

\usepackage{graphicx}

\def\aj{AJ}
\def\apj{ApJ}
\def\apjl{ApJ}
\def\aap{A\&A}
\def\aapr{A\&ARv}

\def\mnras{MNRAS}
\def\degr{$^{\circ}$}
\def\kms{{\rm km~s}$^{-1}$}
\def\Msun{M$_\odot$}
\def\SB9{S$_{\rm B}^9$}

\title{$\delta$ Mus revisited\thanks{Based on data obtained from the ESO Science Archive Facility under request numbers Pourbaix/336063}:Rectifying a 82~yr-old mistake}
\shorttitle{$\delta$ Mus revisited}
\author{Dimitri~Pourbaix,\altaffilmark{1,2}
  and Henri~M.~J.~Boffin,\altaffilmark{3}}
  
  \altaffiltext{1}{Institut d'Astronomie et d'Astrophysique, Universit\'e Libre de Bruxelles, Belgium.}
  \altaffiltext{2}{F.R.S.-FNRS, Belgium.}
  \altaffiltext{3}{ESO, Germany.}
  \fulladdresses{
  \item Boffin: ESO, Karl-Scwharzschild-str. 2, 85748 Garching, Germany (hboffin@eso.org).
    \item Pourbaix: Institut d'Astronomie et d'Astrophysique, CP 226, Universit\'e Libre de Bruxelles, Bld. du Triomphe, 1050 Brussels, Belgium (pourbaix@astro.ulb.ac.be).
    }
\shortauthor{Pourbaix \& Boffin}

\abstract{
  The red giant $\delta$ Muscae is known since 1919 to be a spectroscopic binary, and the first and only orbit was determined in 1936, claiming the period to be 847 days. This was discrepant with the Hipparcos determined astrometric orbit. Using the latest data available for this object -- leading to a 100 yr time-span -- we show here that the correct period is 423 d, and are able for the first time to combine the spectroscopic orbit with the Hipparcos orbit. Using all the available information, we find that the $\sim$1.2~\Msun red giant must have a $\sim 0.3-0.4$~\Msun M dwarf companion, and that the system will soon evolve towards a He WD binary system. Given its relatively short period,  $\delta$ Muscae may be an ideal benchmark for testing astrometric orbits obtained by Gaia for very bright stars.}

\resumen{}

  \addkeyword{astrometry}
  \addkeyword{(stars)binaries: spectroscopic}
  \addkeyword{techniques: radial velocities}

  \SetYear{2018}
  \ReceivedDate{27/04/2018}
  \AcceptedDate{04/06/2018}
\begin{document}
\maketitle

\section{Introduction}

$\delta$ Mus (HIP~63613, HD~112985, HR~4923, ICRS J2000 13:02:16.26474-71:32:55.8752) is a neglected bright K2~III binary star with $V=3.62$ mag and no sign of variability above 4~mmag \citep{Hipparcos}.  It was part of the sample of spectroscopic binaries containing late-type giants that was statistically analysed by \citet{1993A&A...271..125B}. Its spectroscopic orbit, however, has been obtained by \citet{1936ApJ....83..433C} with no revision since its publication, despite the fact that it was classified poor in all versions of the DAO Spectroscopic Binary Catalogues.  In anticipation of the Gaia mission \citep{2016A&A...595A...1G}, \citet{2012A&A...546A..61D} called for an improved estimate of the radial velocity to be used in the HIPPARCOS-Gaia Hundred-Thousand-Proper-Motion project, without any success.  Even though $\delta$ Mus is observed by Gaia (it was missing from Gaia Data Release 1, now present in Gaia Data Release 2 \citep{2018A&A.....GAIADR2}), its radial velocity might still have to come from a ground-based facility (Gaia DR2 does not report any velocity for that object). 

The binary nature of $\delta$ Mus was also noticed by Hipparcos and will likely be noticed by Gaia as well (the single star model was imposed to all the objects present in Gaia DR2, whether ultimately appropriate or not).  The orbital period is short enough to be well covered by Gaia, making this object very well suited for the early validation of the Gaia non-single star processing pipeline, especially as an example of a bright star.  It is therefore mandatory to first validate the ground-based solution which has been considered preliminary for the past 80 years.

The history and status of the only spectroscopic orbit of this system together with an alternative fit are described in Sect.~\ref{sec:SB}, followed by the astrometric counterpart (Sect.~\ref{sec:HIP}).  The  benefit from adding just five radial velocities is the topic of Sect.~\ref{sec:newSB}, while Sect.~\ref{sec:PP} presents the physical properties we can derive for this system.

\begin{figure*}
  \resizebox{0.49\hsize}{!}{\includegraphics{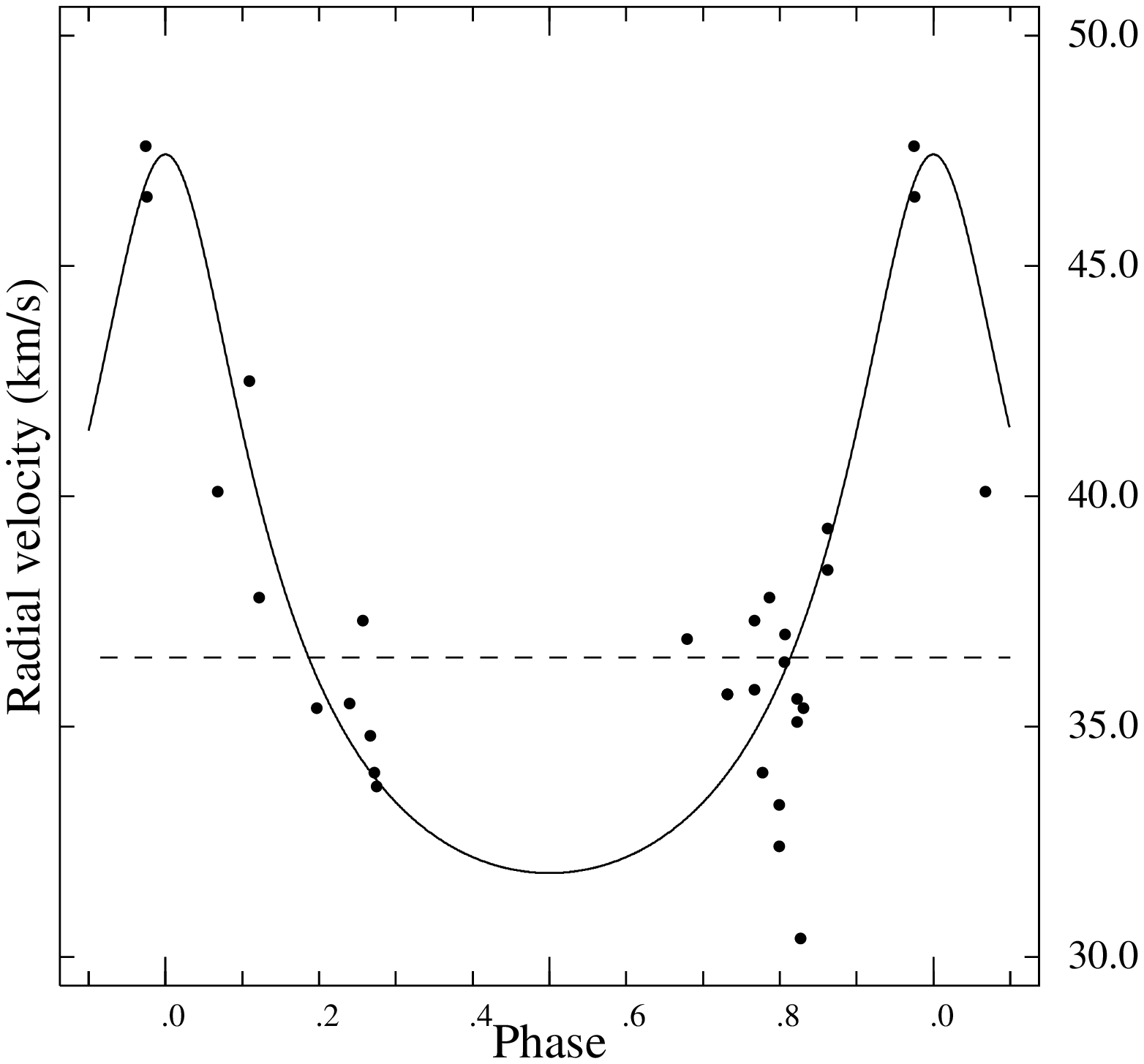}}
  \resizebox{0.49\hsize}{!}{\includegraphics{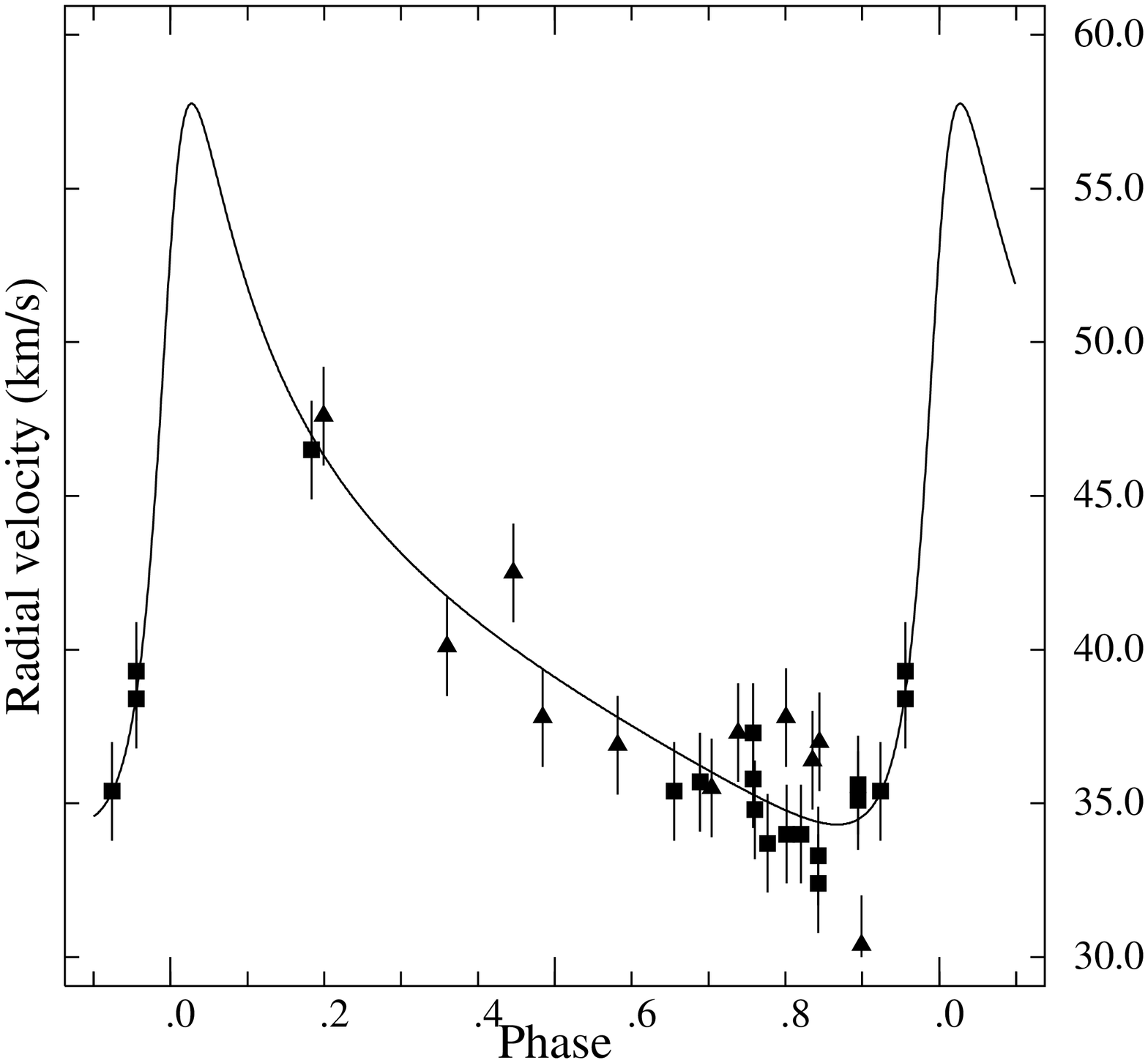}}
  \caption[]{\label{fig:olddata}Original solution by \citet{1936ApJ....83..433C} retrieved from \SB9\ \citep{2004A&A...424..727P} and our alternative solution based on the exact same data (left and right panels, respectively).  In the right panel, the squares denote the Lick data \citep{1928PLicO..16....1C} whereas the triangles are for the Cape ones \citep{1928AnCap..10....8S}.}
\end{figure*}

\section{Spectroscopic orbit}\label{sec:SB}

The first measurement of the radial velocity of $\delta$ Mus dates back to 1904 and it was quickly recognised as variable \citep{1919ApJ....50..161L}.  However, despite that early identification, none of the sets of velocities recorded at the Cape Observatory \citep{1928AnCap..10....8S} and at the Lick Observatory \citep{1928PLicO..16....1C} was accompanied by any tentative orbit.  Finally, \citet{1936ApJ....83..433C} combined the two sets and published the first (and so far unique) spectroscopic orbit (847-day period) of that system.  Despite the orbit being described as {\em preliminary} by the author himself, it has not been confirmed by any independent study for the past 80 years.  Actually, only a handful of new radial velocities have been reported since 1928 \citep{1980AJ.....85..858S,2014A&A...561A.126D}.

With the exact same radial velocities as \citet{1936ApJ....83..433C}, it is nevertheless possible to derive a rather different solution, in particular for the period which is essentially one half of the original estimate.  That very ill-defined solution is listed in Tab.~\ref{TAB:OLDSBORBIT} together with \citeauthor{1936ApJ....83..433C}'s orbit.  The two solutions are plotted in Fig.~\ref{fig:olddata}.  The reason why our alternative orbit is so poorly constrained is clear from the plot: with the new period, there is no data covering the upper part of the curve.

\begin{table}[htb]
  \caption[]{\label{TAB:OLDSBORBIT}Two alternative spectroscopic orbits based on the original spectroscopic data (Spencer Jones, 1928; Campbell, 1928)}
  \centering
  \begin{tabular}{lcc}\toprule
    Parameter & \citet{1936ApJ....83..433C} & Alternative solution \\ \midrule
    $P$ (days) & 847 & $424.0\pm1.2$\\
    $T_0$ (JD-2\,400\,000) & 21790 & $16597\pm22$\\
    $e$ & 0.4 & $0.60\pm0.50$\\
    $\omega$ (\degr) & 0 & $126\pm51$\\
    $K_1$ (\kms) & 7.8 & $12.0\pm9.7$\\
    $V_0$ (\kms) & +36.5 & $+41.0\pm3.4$\\
    \bottomrule
    \end{tabular}
\end{table}

The original papers did not report the uncertainties on the observations.  The residuals derived from our solution have a zero mean and a standard deviation of 1.6 \kms.  Assuming that value as uncertainty for all the velocities, the goodness of fit, $F2$, can be used to assess the quality of the fits. It is defined as
\[
F2=\sqrt{\frac{9\nu}{2}}(\sqrt[3]{\frac{\chi^2}{\nu}}+\frac{2}{9\nu}-1)
\]
where $\nu$ is the number of degrees of freedom and $\chi^2$ is the weighted sum of the squares of the differences between the predicted and the observed positions \citep{FuAs}. $F2$ \citep[Wilson \& Hilferty's cube root transformation,][]{Wilson-1931} follows a N(0, 1) distribution \citep{KeAdThSt1}.  For \citeauthor{1936ApJ....83..433C}'s orbit and ours, the $F2$ are respectively 3.66 and 0.78.  In terms of fit only, our alternative solution fits the data much better than the original one.

\section{Astrometric observations}\label{sec:HIP}

Like every other objects in that magnitude range, $\delta$ Mus was observed by Hipparcos \citep{Hipparcos}.  However, in this particular case, the binary nature of the source was also noticed and the full-fledged orbital model was even necessary to fit the observations.  Despite the availability of a spectroscopic orbit \citep{1936ApJ....83..433C}, the seven parameters of the astrometric orbit were fitted independently.

\begin{table}[htb]
  \caption[]{\label{tab:Astroorb}Astrometric orbit from Hipparcos (ESA, 1997).}
  \centering
  \begin{tabular}{lc}\toprule
    Parameter & Value \\ \midrule
    $P$ (days) & $422.0266\pm5.3908$ \\
    $T_0$ (JD-2\,400\,000) & $47947.6687\pm27.5934$ \\
    $e$ & $0.4918\pm0.1241$ \\
    $\omega$ (\degr) & $316.39\pm26.13$ \\
    $\Omega$ (\degr) & $59.27\pm5.29$ \\
    $i$ (\degr) & $120.07\pm6.02$\\
    $a_0$ (mas) & $11.67\pm1.02$ \\
    \bottomrule
    \end{tabular}
\end{table}

Even though the astrometric period and our revised spectroscopic one look similar, the consistency of the eccentricities comes from their rather large uncertainties only and the arguments of the periastron ($\omega$) are totally discrepant.  \citet{2003A&A...398.1163P} showed that no satisfactory astrometric fit could be based on \citeauthor{1936ApJ....83..433C}'s orbit, so if one looks for an improved consistency between spectroscopy and astrometry, it should come from the radial velocities.

\begin{table}
  \caption[]{Radial Velocity measurements of $\delta$ Mus.  The phases and residuals are derived from the orbit in Tab.~\ref{TAB:NEWSBORBIT}.}\label{tab:RVs}
  \centering
\begin{tabular}{lccccc}
\toprule
Epoch	& Phase & Rad.~Vel.	& uncertainty & Residual&  Ref.\\
JD-2,400,000.0	& &(km/s)	&(km/s)	& (km/s) &	\\
\midrule
16519.79    & 0.726 & 34.00 & 1.6  &  -0.96& 2\\
16564.79    & 0.833 & 35.40 & 1.6  &  +0.10& 2\\
16874.82    & 0.565 & 35.40 & 1.6  &  -0.84& 2\\
16938.62    & 0.716 & 34.00 & 1.6  &  -1.02& 2\\
18380.49    & 0.124 & 47.60 & 1.6  &  +0.12& 1\\
19078.79    & 0.774 & 32.40 & 1.6  &  -1.59& 2\\
19078.79    & 0.774 & 33.30 & 1.6  &  -2.49& 2\\
19098.83    & 0.821 & 35.10 & 1.6  &  +0.43& 2\\
19098.83    & 0.821 & 35.60 & 1.6  &  -0.07& 2\\
19949.35    & 0.831 & 30.40 & 1.6  &  -4.91& 1\\
20199.62    & 0.422 & 37.80 & 1.6  &  -0.42& 1\\
21769.54    & 0.132 & 46.50 & 1.6  &  -0.39& 2\\
22015.72    & 0.714 & 34.80 & 1.6  &  -0.23& 2\\
22022.72    & 0.731 & 33.70 & 1.6  &  -1.26& 2\\
22409.84    & 0.646 & 35.70 & 1.6  &  +0.21& 2\\
22409.84    & 0.646 & 35.70 & 1.6  &  +0.21& 2\\
22439.74    & 0.716 & 35.80 & 1.6  &  +2.27& 2\\
22439.74    & 0.716 & 37.30 & 1.6  &  +0.77& 2\\
22456.44    & 0.756 & 37.80 & 1.6  &  +2.90& 1\\
22472.40    & 0.793 & 36.40 & 1.6  &  +1.46& 1\\
22473.38    & 0.796 & 37.00 & 1.6  &  +2.02& 1\\
22520.54    & 0.907 & 38.40 & 1.6  &  +1.20& 2\\
22520.54    & 0.907 & 39.30 & 1.6  &  +0.30& 2\\
22694.65    & 0.319 & 40.10 & 1.6  &  -0.16& 1\\
22729.58    & 0.401 & 42.50 & 1.6  &  +3.97& 1\\
22840.33    & 0.663 & 35.50 & 1.6  &  +0.16& 1\\
22854.39    & 0.696 & 37.30 & 1.6  &  +2.19& 1\\
23212.39    & 0.542 & 36.90 & 1.6  &  +0.38& 1\\
43969.71    & 0.595 & 33.30 & 2.6  &  -2.64& 3\\
43970.70    & 0.598 & 34.20 & 2.4  &  -1.71& 3\\
48372.721   & 0.000 & 50.48 & 0.27 &  -0.01& 4\\
49766.882   & 0.295 & 40.70 & 0.26 &  -0.08& 4\\
53070.59492 & 0.102 & 48.90 & 0.56 &  +0.11& This work\\
\bottomrule
\end{tabular}

1: \citet{{1928AnCap..10....8S}}; 2: \citet{1928PLicO..16....1C}; 3: \citet{1980AJ.....85..858S}; 4: Coravel (Udry S., Priv. Comm., 2001)
\end{table}

\section{Just five more points}\label{sec:newSB}
Whereas four measurements of the radial velocities have been reported since 1928, only two are public \citep{1980AJ.....85..858S}.  Obtained at Cerro Tololo, they date back to 1979 but were taken just one day apart and would thus likely count as one epoch in any orbital fit.  The remaining two radial velocities were measured with Coravel South in the framework of obtaining the radial velocity of the Hipparcos stars.  The average of the two observations was published by \citet{2014A&A...561A.126D} but the individual measurements were kindly supplied (S.~Udry, private communication) for a previous investigation \citep{2003A&A...398.1163P}.  A fifth radial velocity was derived from a public FEROS spectrum, retrieved from the ESO Science Archive (Prog. Id. 072.D-0235(B)). The FEROS spectrum, obtained on 6 March 2004, had an exposure time of 25 s and a signal-to-noise ratio of about 70, at a spectral resolution of 48,000. The radial velocity was derived by cross-correlating the spectrum with a synthetic spectrum of a 4500 K, $\log g$=2.5 star.

\begin{figure}[htb]
  \centering
  \resizebox{\hsize}{!}{\includegraphics{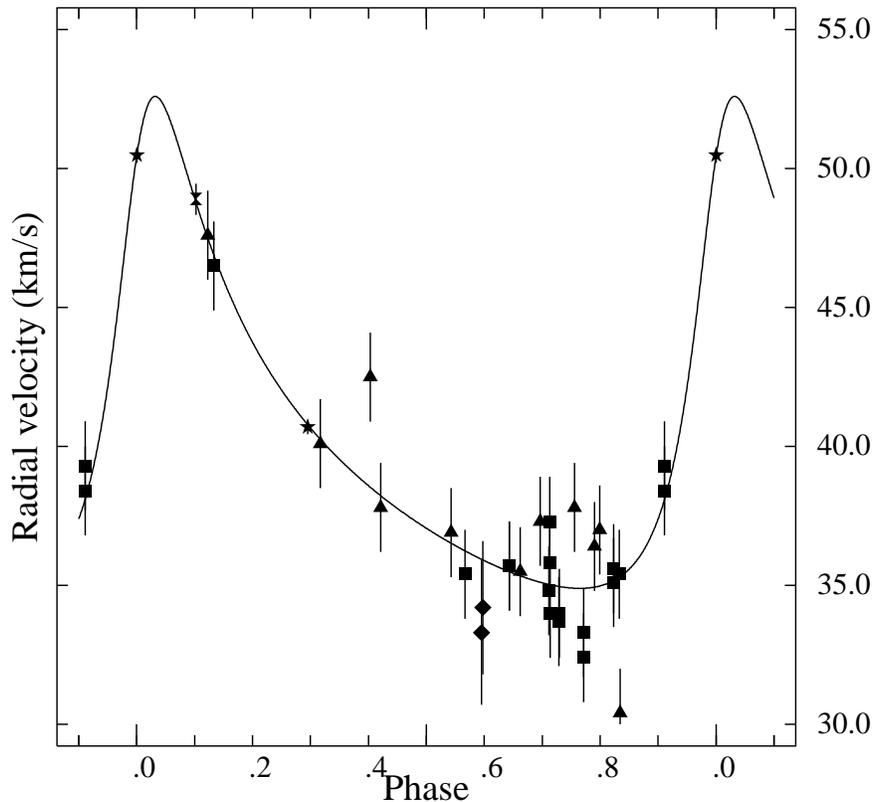}}
  \caption{New, improved orbit for $\delta$ Mus, based on the elements in Table~\ref{TAB:NEWSBORBIT}. The squares denote the Lick data \citep{1928PLicO..16....1C}, the triangles are the Cape ones \citep{1928AnCap..10....8S}, the diamonds are the Cerro Tololo velocities \citep{1980AJ.....85..858S}, the stars are the Coravel South data (S.~Udry, priv. comm.), and the diabolo denotes the FEROS based velocity (this work).}\label{fig:newSBorbit} 
\end{figure}

With these five points added to the set originally used by \citet{1936ApJ....83..433C}, the baseline grows from 18 to exactly 100 years.  Owing to the poor precision of the original radial velocities (about 1.6~\kms) and even the Cerro Tololo ones ($\sim 2.6$~\kms) with respect to the Coravel ones ($\sim 0.26$~\kms) and the latest FEROS one ($\sim 0.56$~\kms), the likely discrepancy between the velocity zero points of the five observatories is neglected.  For completeness, the radial velocity measurements of $\delta$ Mus are provided in Table~\ref{tab:RVs}.  The adopted weight is the reciprocal of the square of the uncertainty. The resulting orbit is listed in Table \ref{TAB:NEWSBORBIT} and plotted in Fig.~\ref{fig:newSBorbit}.  That orbit is characterised by an $F2$ of 0.87, slightly larger than with our previous solution but still within the boundaries of a good fit.

\begin{table}[htb]
  \caption{Spectroscopic orbit based on all the available data.}\label{TAB:NEWSBORBIT}
  \centering
  \begin{tabular}{lc}\toprule
    Parameter & Revised solution \\ \midrule
    $P$ (days) & $423.2\pm0.10$\\
    $T_0$ (JD-2\,400\,000) & $25945\pm5.5$\\
    $e$ & $0.52\pm0.06$\\
    $\omega$ (\degr) & $319\pm4.2$\\
    $K_1$ (\kms) & $8.8\pm0.38$\\
    $V_0$ (\kms) & $+40.2\pm0.20$\\
    \bottomrule
    \end{tabular}
\end{table}

The agreement between the spectroscopic and astrometric solutions is now much better and yet the two fits are still independent.  Even though there are twice as many astrometric observations as spectroscopic ones, the former cover 2.6 orbital periods only, thus making the spectroscopic solution much more precise (and hopefully accurate).  Assuming it in the astrometric one would reduce the number of fitted parameters and therefore increase their precision.  There are essentially two ways of using a spectroscopic orbit in an astrometric orbital fit and comparing the results of the two approaches makes it possible to assess reliability of both astrometric fits \citep{2001A&A...369L..22P,2001A&A...372..935P}.

Plugging the elements from Tab.~\ref{TAB:NEWSBORBIT} into such a fit now leads to very consistent results: all the statistical indicators prescribed by \citet{2001A&A...372..935P} are satisfied.  Constraining the astrometric fit with $\omega$, $e$, $P$, $T_0$, and even $K_1$ gives an astrometric solution whose error bars are substantially reduced as fewer elements are fitted (Tab.~\ref{tab:newAstroOrb}).

\begin{table}[htb]
  \caption[]{Revised astrometric orbit based on the Hipparcos (ESA, 1997) data and the spectroscopic solution from Tab.~\ref{TAB:NEWSBORBIT}.  $a_0$ denotes the semi-major axis of the absolute orbit of the photocentre of the system.}\label{tab:newAstroOrb}
  \centering
  \begin{tabular}{lc}\toprule
    Parameter & Value \\ \midrule
    $\Omega$ (\degr) & $58\pm2.7$ \\
    $i$ (\degr) & $120\pm2.3$\\
    $a_0$ (mas) & $12.1\pm0.35$ \\
    $\varpi$ (mas) & $35.6\pm0.56$\\
    $\mu_{\alpha*}$ (mas yr$^{-1}$) & $263.6\pm0.48$\\
    $\mu_{\delta}$ (mas yr$^{-1}$) & $-23.4\pm0.48$\\
    \bottomrule
    \end{tabular}
\end{table}

\begin{figure}[htb]
  \centering
  \resizebox{\hsize}{!}{\includegraphics{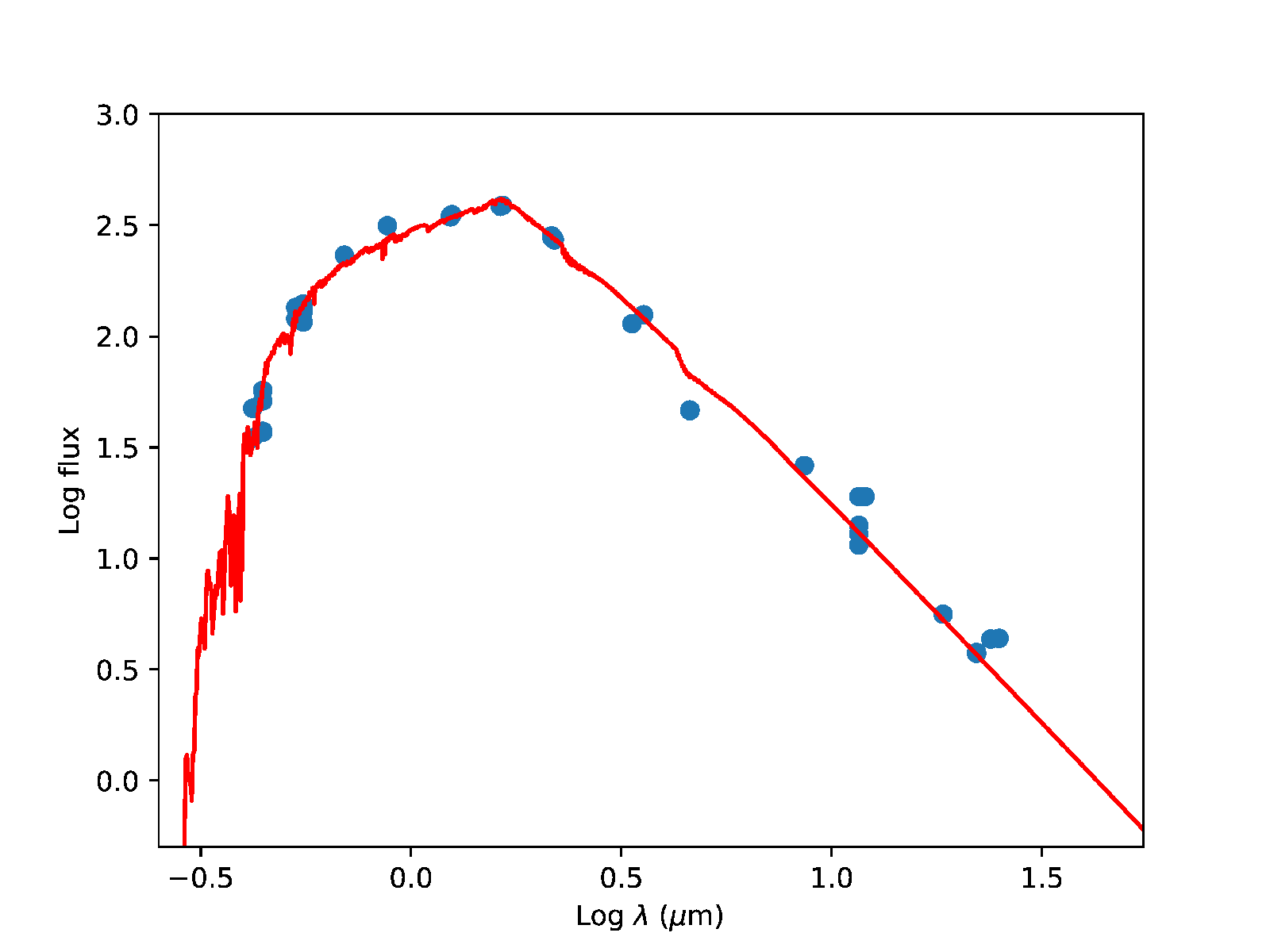}}
  \caption{Comparison between the photometry of $\delta$ Mus as retrieved from Vizier and a model for a $T_{\rm eff}$=4500 K, $\log g=2.5$ from \citet{2004astro.ph..5087C}.}\label{fig:SED}
\end{figure}

\section{Physical properties}
\label{sec:PP}

2MASS \citep{2003tmc..book.....C} reports a magnitude $K=0.946\pm0.282$ which, for a red giant, translates to $T_{\rm eff}=4500$K and a bolometric correction in $K$ of 2.3 \citep{1998A&A...337..321B}.  We have confirmed this value, by downloading the photometry of $\delta$ Mus over the widest spectral range from CDS/Simbad, using Vizier and compared it to a solar-metallicity model of a $T_{\rm eff}$=4500 K, $\log g=2.5$ star \citep[][\footnote{http://wwwuser.oats.inaf.it/castelli/grids.html}]{2004astro.ph..5087C}. A simple rescaling provides a perfect match between observations and model (Fig.~\ref{fig:SED}).  This also indicates that the companion does not produce any noticeable flux in the wavelength range from 420 nm to 60 $\mu$m.  

\begin{figure}[htb]
  \centering
  \resizebox{\hsize}{!}{\includegraphics{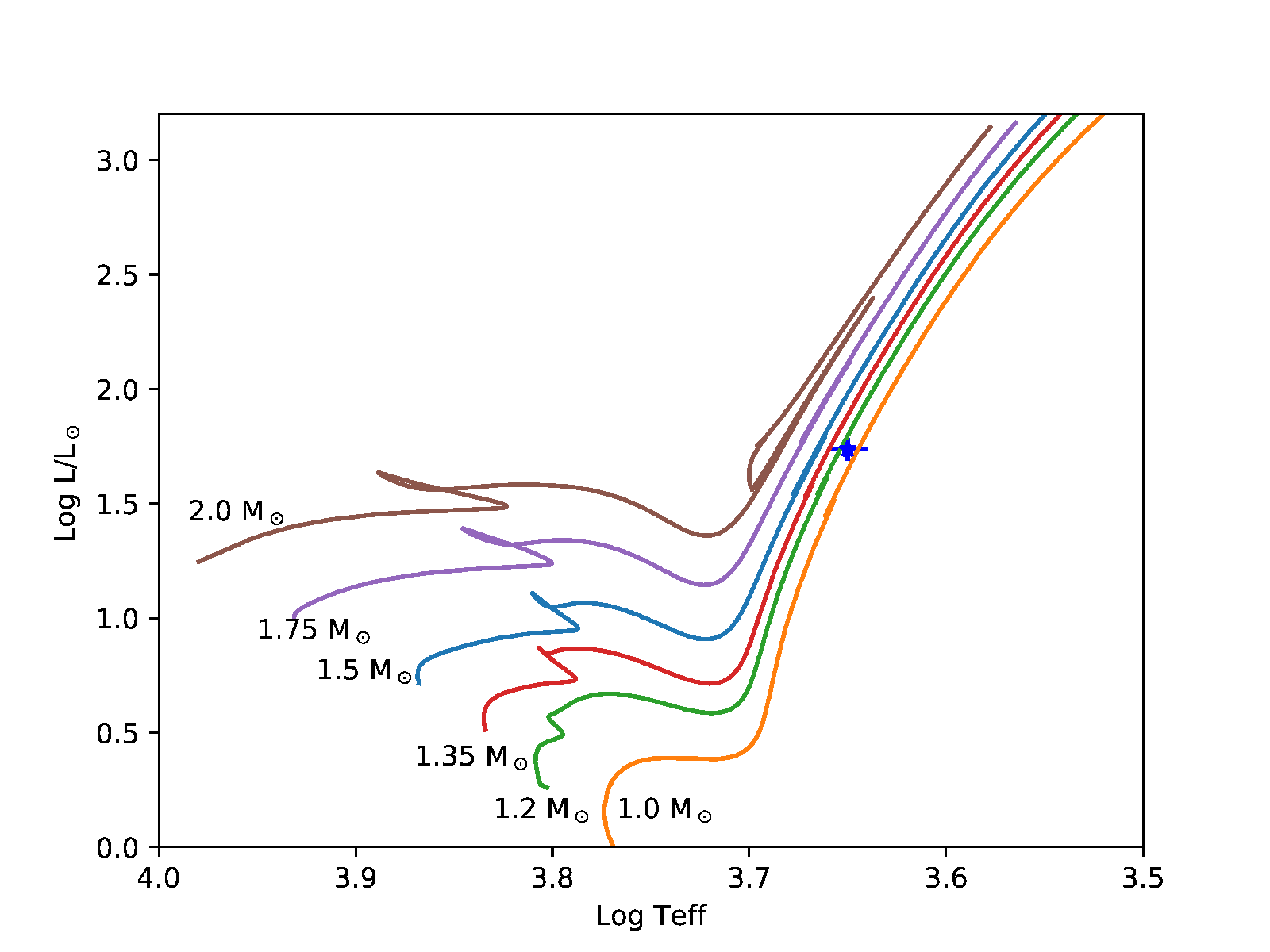}}\\
  \resizebox{\hsize}{!}{\includegraphics{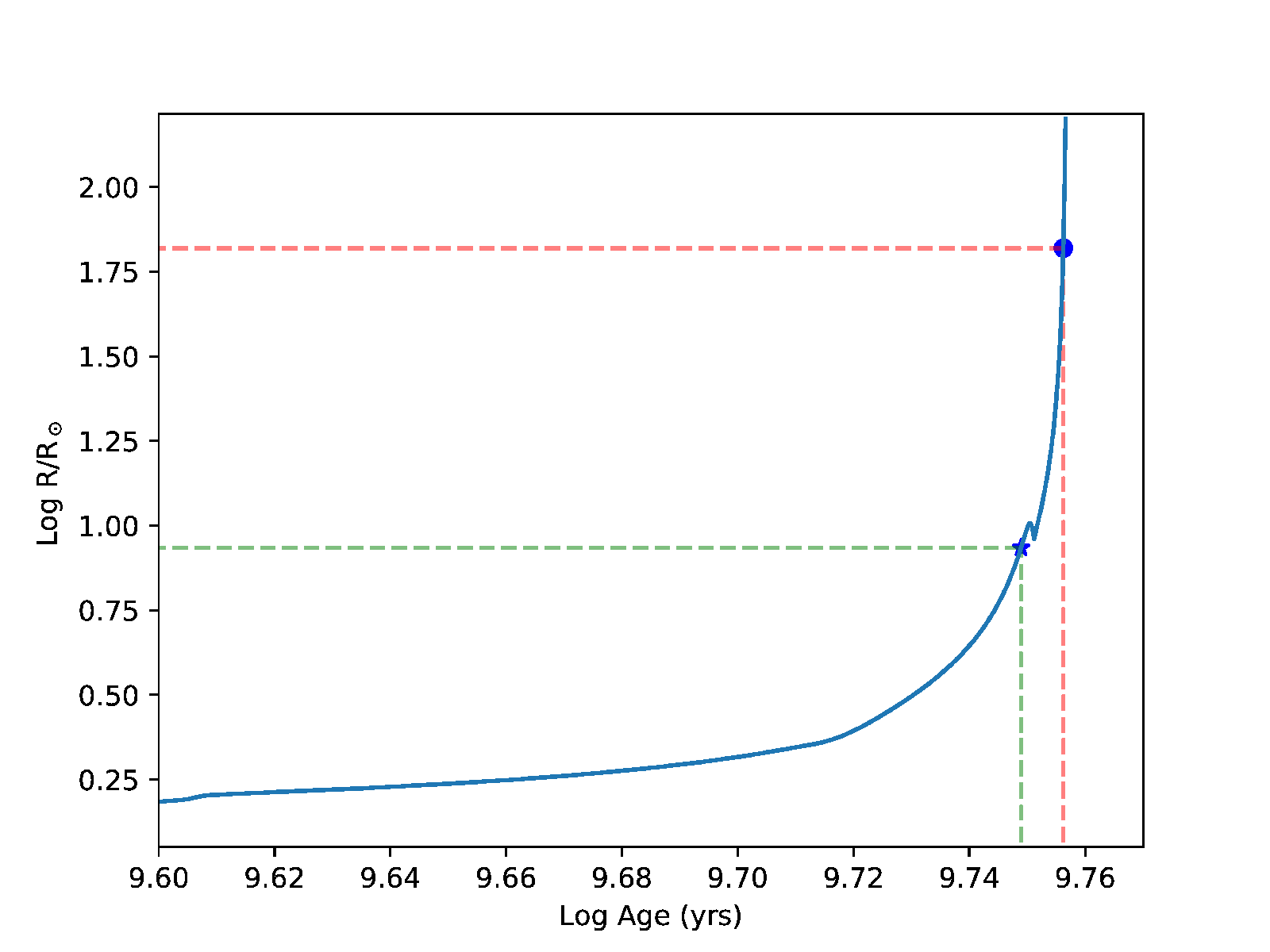}}
  \caption{(top) Hertzsprung-Russell diagram with the position of $\delta$ Mus indicated with a star, compared to evolutionary tracks from \citet{1994A+AS..106..275B}. (bottom) Evolution of the radius of a 1.2~\Msun stellar model from  \citet{1994A+AS..106..275B}. The current position of $\delta$ Mus is indicated with a star, while the value of 66~R$_\odot$ that would lead to Roche-lobe overflow (see text) is shown as a dot.}\label{fig:HR}
\end{figure}

Such a temperature is associated to a star with a mass close to 1.5~\Msun\ \citep{2016MNRAS.456.3655M}.  This can be refined by placing the star in an Hertzsprung-Russell diagram. 
Using our revised Hipparcos parallax ($\varpi=35.6\pm0.56$~mas), we can derive a bolometric luminosity of $M_{\rm bol}= 1.00$, or $L/L_\odot= 31.24$. With the given temperature, we obtain a radius of the giant of 8.6R$_\odot$. 
Assuming a solar metallicity \citep{1993AJ....106...80E}, the evolutionary tracks of \citet{1994A+AS..106..275B} bracket the mass in the range $1.05 - 1.35$~\Msun\, (Fig.~\ref{fig:HR}).  

The orbit from Tab.~\ref{TAB:NEWSBORBIT} results in a spectroscopic mass function, $f(m)$, of $0.0190\pm0.0034$~\Msun.  Combined with the mass of the primary and the inclination from the astrometry (Tab.~\ref{tab:newAstroOrb}), this yields a $\sim0.4$~\Msun\ secondary.  With such a mass, it is either a  $\sim$M3 dwarf \citep{Torres-2010} or an Helium white dwarf \citep{2007MNRAS.375.1315K}, in which case it would have to be the result of some binary interaction that cut the normal stellar evolution. This would be quite surprising as the eccentricity of the system is quite large, and not in agreement with the circular orbit we would expect for such a post-mass transfer system -- in fact the eccentricity is well-above the envelope found for post-mass transfer red giants \citep{2017A&A...597A..68V}.   \citet{2002A&A...383..188M} list $\delta$ Mus as a reference star for their atlas of symbiotic stars, and indeed its spectrum exhibits no emission line, nor any sign of accretion -- a white dwarf companion is therefore not detected.  Moreover, the star is not known for any variability, that could betray some accretion onto a compact object. We conclude therefore that a main-sequence M star is much more likely.

From the mass of both components, the period and the parallax, the semi-major axis of the relative orbit turns out to be about 46 mas, which can be resolved with interferometric facilities.  If the secondary is a M dwarf, the difference is of the order of 9 magnitudes in $V$ and even less in $K$ --  one cannot therefore exclude the possibility of resolving the two components with the GRAVITY instrument (General Relativity Analysis via VLT Interferometry) attached to the Very Large Telescope Interferometer \citep[VLTI;][]{2017A&A...602A..94G}, and we encourage the community to do so. Given the brightness of $\delta$ Mus, this wouldn't require much observing time.  We note that the angular diameter of $\delta $ Mus would be $\sim3$ milli-arcseconds and would thus relatively easily be resolved by the VLTI. 

The semi-major axis of the relative orbit ($a$) and the semi-major axis of the absolute orbit of the photocenter ($a_0$, Tab.~\ref{tab:newAstroOrb}) are linked through the fractional mass and the fractional luminosity of the components:
\begin{eqnarray*}
a_0&=&\left(\frac{M_2}{M_1+M_2}-\frac{L_2}{L_1+L_2}\right)a\\
   &=&\left(\frac{M_2}{M_1+M_2}-\frac{L_2}{L_1+L_2}\right)\sqrt[3]{(M_1+M_2)P^2}
\end{eqnarray*}
This yields a fractional luminosity consistent with 0.  That does not come as a surprise as the dynamic range of Hipparcos does not exceed 4 mag (whereas we expect the magnitude difference to be of the order of 9 mag).  One can therefore assume that the photocenter is the primary and, indeed, the astrometry based amplitude of the radial velocity curve ($K_{1,{\rm astr}}= 9.8\pm0.8$\kms) matches the spectroscopic estimate (Tab.~\ref{TAB:NEWSBORBIT}).

The semi-major axis of the relative orbit is 1.29 au and, with the current parameters, the Roche lobe radius of the primary at periastron is about 66 R$_\odot$, a value much larger than the current radius, and the primary is therefore well inside its Roche lobe. According to the stellar evolutionary tracks, however, in about 95 Myr, the giant star will have reached such a radius and Roche lobe overflow will occur. As the mass ratio is well above unity (about 3.4 for now), this mass transfer will be dynamically unstable, leading to the formation of a Common Envelope and the spiral-in of the companion. The end product will be a short period (less than a few days) binary containing a $\sim 0.38$~\Msun He WD (the Helium core mass at the start of the mass transfer) and a M dwarf. Thus $\delta$ Mus is a nice example of a progenitor of He WD binaries.

\section{Conclusion}

Using the latest available spectroscopic data, we have established that the orbital period of the system containing $\delta$ Mus is 423 days, confirming the Hipparcos result and showing that the community has been lured by an alias for the past 80 years.  

Alhough $\delta$ Mus has already been observed by Gaia, a first tentative astrometric orbit will come out with the third data release anticipated for late 2020.  Observations with GRAVITY could be used to further constraint the astrometric orbit, and in the case Gaia can observe it, provide a useful test bench for the handling of very bright stars with Gaia. 
 
The orbit of $\delta$ Mus from \citet{1936ApJ....83..433C} was graded 1 (i.e ``worst'') in \SB9\ \citep{2004A&A...424..727P}, and for good reasons apparently. We note that there are 287 more orbits (for 282 systems) sharing the same poor grade in \SB9\.  Among these systems, only 45 have had their orbit improved later on.  Clearly some orbits of these 237 systems may turn out to be wrong as well and we urge readers to exert the uttermost caution when using such orbits for further analysis (statistical or others).

\section*{Acknowledgements}
The authors thank the referee for his/her careful reading.  This research has made use of the Simbad and VizieR data bases, operating at CDS, Strasbourg, France.

\end{document}